\documentclass[aps, prb, reprint, superscriptaddress,  
               showpacs, floatfix, groupedaddress]{revtex4-1}
\usepackage{graphicx}
\usepackage{amsmath}
\usepackage{amssymb}
\usepackage{soul}
\usepackage{ulem}
\usepackage[dvipsnames]{xcolor}
\usepackage{xspace}

\begin{document}

\title{Quantum transport in graphene Hall bars - effects of vacancy 
       disorder} 

\author{M. D. Petrovi\'c} \email{marko.petrovic@uantwerpen.be}   
\author{F. M. Peeters} \email{francois.peeters@uantwerpen.be}
\affiliation{Department of Physics, University of Antwerp, \\ 
             Groenenborgerlaan 171, B-2020 Antwerp, Belgium}

\begin{abstract}
Using the tight-binding model, we investigate the influence of vacancy
disorder on electrical transport in graphene Hall bars in the presence of
quantizing magnetic fields. Disorder, induced by a random distribution of
 monovacancies, breaks the graphene sublattice symmetry
and creates states localized on the vacancies. These states are observable in
the bend resistance, as well as in the total {DOS}. Their energy is proportional
to the square root of the magnetic field, while their localization length is
proportional to the cyclotron radius. At the energies of these localized
states, the electron current flows around 
the monovacancies and, as we show, it can follow unexpected paths depending
on the particular arrangement of vacancies. We study how these localized
states change with the vacancy concentration, and what are the effects of
including the next nearest neighbor hopping term. Our results are also
compared with the situation when double vacancies are present in the system.
Double vacancies also induce localized states, but their energy and
magnetic field dependences are different. Their localization energy scales
linearly with the magnetic field, and their localization length appears not to
depend on the field strength.
\end{abstract}

\maketitle{}

\section{Introduction} 
The discovery of graphene,~\cite{novoselov_graphene} a material with a linear
low-energy spectrum, generated new interest in the quantum Hall effects
governed by relativistic particles. Unusual quantum Hall resistance plateaus
were observed in graphene.~\cite{graphene_qhe_zhang, novoselov_qhe} Later
experiments reported new, more detailed, features such as the splitting of the
zeroth Landau level (LL) due to breaking of the valley and spin 
degeneracies.~\cite{zhang.prl.2006, li.prl.2009}

Disorder in experimentally available honeycomb graphene lattices is
inevitable, whether it is structural like reconstructed and non-reconstructed
vacancies, substituted carbon atoms, or it originates from charged impurities
such as adatoms. Therefore, disorder in graphene is a very active area of
research, both experimentally, in devising ways to characterize
it,~\cite{ugeda_prl_2010} and theoretically in studying its influence on
electron transport,~\cite{duffy, tijerina} with even possible applications in
future spintronic devices.~\cite{droth_prb_2015} Due to the relativistic
nature of its charge carriers, disordered graphene offers a tabletop
environment for the study of previously experimentally unaccessible phenomena,
such as for example the atomic collapse reported recently in charged vacancies
in graphene.~\cite{dean_nature} Vacancy disorder in case of a zero external
magnetic field was extensively studied in Refs.~\onlinecite{pereira_prb,
pereira_prl, cresti_new}, where new states localized around missing carbon
atoms were reported. Effects of vacancies in the quantum Hall regime were
studied in Refs.~\onlinecite{leconte_prb_2016, ortman_prl_2013}, which
reported on the occurrence of a zero-resistance quantum Hall plateau, and
breaking of the Landau level degeneracy. Graphene with on-site potential
disorder was also used in Ref.~\onlinecite{garcia_prl_2015} to test a new
numerical approach to calculate the Kubo conductivities.

In this paper, we simulate the transport of electrons in a Hall bar made from
a single layer of graphene. Our main goal is to study the influence of various
types of vacancy disorder on the electron transport in the quantum Hall
regime. We report that vacancy disorder can cause the appearance of new states
in the Landau spectrum, which are observable in the bend resistance, as well
as in the total density of states (DOS) and in the distributions of
eigenenergies in a closed system (a system detached from the leads). For
monovacancies, the energies of these new states scale as the square root of
the magnetic field, similarly to the energies of relativistic Landau levels,
but with a different scaling coefficient. The local density of states (LDOS)
reveals a strong localization around the monovacancy sites, with localization
length proportional to the cyclotron radius.  The localization on divacancies
is somewhat different: their localization energy scales linearly with the
magnetic field, while their localization radius appears to be constant. We
further study how the electron current flows in the presence of vacancies, and
what are the effects of the next nearest neighbor interaction (NNN).

This paper is organized as follows: In section~\ref{s:sys} we describe our
system and methods used to obtain our results. In order to focus on specific
aspects of the problem, ranging from vacancy concentration to NNN hopping, we
discuss our results (Sec.~\ref{s:res}) in several subsections
(from~\ref{ssDisorder} to~\ref{ssNNN}). All these insights are combined and
summarized in the last, concluding section (Sec.~\ref{s:end}).

\section{System and Methods \label{s:sys}}
\begin{figure}[t]
\includegraphics{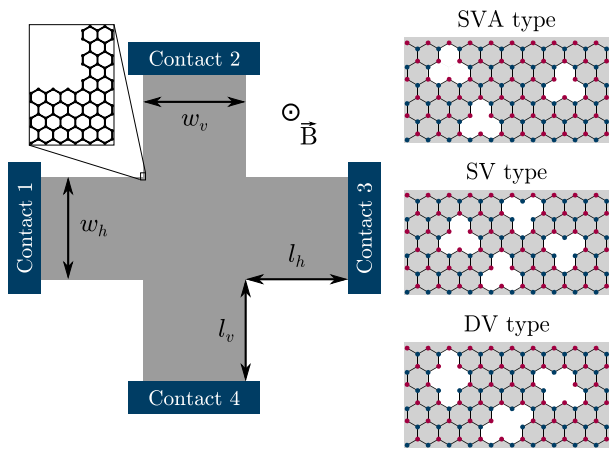}
\caption{\label{f:sys}
    (Color online) Graphene Hall bar system (left), and three studied 
    disorder types (right). Widths of the horizontal and vertical
    leads are set to $w_h$ = 49.71 nm, and $w_v$ = 49.94 nm, respectively,
    while lengths of the horizontal and vertical arms ($l_v$ and $l_h$) are
    equal, and set to 50 nm. Before disorder is introduced, all
    edges are considered to be clean, meaning that there are no dangling
    bonds on them.} 
\end{figure}

The studied system is shown in Fig.~\ref{f:sys}, it is a graphene Hall bar
with zigzag edges along the horizontal leads, and armchair edges along the 
vertical leads. The width of the vertical, armchair, arms ($w_v$) is 
chosen so that the corresponding leads are metallic, meaning that there is 
no gap around  zero energy.

We introduce vacancy disorder in this system by randomly removing carbon
atoms from the graphene lattice. Three different disorder types are studied,
as shown in the right insets in Fig.~\ref{f:sys}. The first is a single
vacancy/single sublattice disorder (SVA). Here, we randomly remove carbon
atoms only from one  sublattice (e.g.~sublattice A). 
The second type is
ordinary single vacancy disorder (SV), where carbon atoms are removed
without any respect to the sublattice to which they belong. The third type is
a double vacancy disorder (DV), where only pairs of neighbouring atoms, each
belonging to a different sublattice, are removed. It is known
that single vacancies (or monovacancies) break the sublattice symmetry, while 
divacancies preserve it. Here we choose two types of monovacancy
distributions, since one (SV) should preserve the sublattice symmetry on the
average, while the other (SVA) is the extreme case of sublattice symmetry
breaking. 

When discussing the effects of vacancy disorder, it is important to
investigate the general effects introduced by disorder, and to separate them
from  effects that occur only for some specific disorder distributions.
Therefore, we  will present two types of results. In order to capture the
general disorder effects, for each disorder type and concentration, we perform
calculations over a sample of $N = 10$ different vacancy distributions.
Results for specific distributions $R_i$ ($i= 1, 2, \ldots, 10$) are then
averaged $\bar{R} = \sum_{i=1}^{N} R_i / N$ (we mark the averaged quantities
with a bar line on top). On the other hand, in order to better understand the
origin of these effects, we often analyze results for some specific
distribution, or compare results of several different distributions.

For our numerical calculations  we use KWANT, a software package designed  to
simulate electron transport in the quantum regime.\cite{kwant} KWANT is based
on the so-called wave function formulation of the scattering problem, a method
which is mathematically equivalent to the non-equilibrium Green's function
method, but according to Ref.~\onlinecite{kwant} it is numerically more
stable. We define graphene material in KWANT using the tight-binding model
Hamiltonian 
\begin{eqnarray}
\hat{\mathbf{H}} & = & \sum_{\langle i, j \rangle} 
             (\tilde{t}_{ij} \hat{c}_{i}^\dagger \hat{c}_{j} + H.c.) + 
             \sum_{\langle\langle i, k\rangle\rangle}
             \left(\tilde{t}'_{ik}
                 \hat{c}_{i}^\dagger \hat{c}_{k} + H.c.\right)  
\end{eqnarray}
where $\hat{c}^\dagger_i$ ($\hat{c}_i$) create (annihilate) a $p_z$ electron
on the $i$-th carbon atom.  No external electric potential is included, except
that of the back gate which controls the Fermi energy. The hopping terms
$\tilde{t}_{ij} = t e^{i\varphi_{ij}}$ and $\tilde{t}_{ik}' = t'
e^{i\varphi_{ik}}$ are defined using the electron nearest neighbor hopping
energy $t = -2.7$ eV, the NNN hopping term $t'$, and the Peierls phase factor
$\varphi_{ij}$ (which we discuss below). Although most of our results deal
only with  nearest neighbor interaction \mbox{($t'= 0$)}, in the last
subsection of the next part (\ref{ssNNN}) we comment on the effects of a
nonzero NNN term.

Defining a magnetic field in a multi-lead system, where some leads point in
different directions, is a problem that needs to be carefully considered.
Vector potential along the leads needs to be translatory invariant in order to
simulate each lead as a semi-infinite system.  Following this condition, we
set the vector potential in horizontal leads using the Landau gauge
$\vec{A}_H=-By\vec{e}_x$, and that in vertical leads  as
$\vec{A}_V=Bx\vec{e}_y$. To connect these two, the gauge in the main region is
set to change smoothly from $\vec{A}_H$ to $\vec{A}_V$ in the upper and lower
arms of the cross. This is achieved by using an additional scalar function
$f(x, y)$ which rotates the vector potential $\vec{A}'=\vec{A} + \vec{\nabla}
f$ locally, without changing the orientation and strength of the magnetic
field.  This scalar function is defined in Ref.~\onlinecite{baranger_bf} as 

\begin{equation}
 f(x, y) = Bxy\,{\sin^2}\theta + 
           \frac{1}{4}B\left(x^2 - y^2\right)\sin2\theta,
\end{equation}
where $\theta$ is the angle of rotation. In order to apply $f(x, y)$
only in a specific subregion of the cross, we multiply it with a smooth step
function \mbox{$\xi_i(y)=\frac{1}{2}\left[1 + \tanh(2(y-y_0)/d) \right]$},
which is nonzero only very close to one of the vertical leads (here the index
$i$ specifies the lead number). Previous expression defines $y_0$ as a
crossover position, where $\xi_i(y_0)=\frac{1}{2}$, and $d$ as a  width of the
crossover region, where $\xi_i$ smoothly goes from 0 to 1. For our numerical 
calculations, we used $d = l_v/5 = 10$ nm. Based on this, we can define a 
rotation function for the second lead  

\begin{eqnarray}
F_{2}(x, y) & = & f(x, y)\xi_{2}(y) \nonumber \\ 
      & = & \frac{1}{2}Bxy \left[1 + \tanh\left(2\frac{y-y_u}{d}\right) \right],
\end{eqnarray}
and similarly for the  fourth lead
\begin{eqnarray}
F_{4}(x, y) & = & f(x, y)\xi_4(y) \nonumber \\
            & = & \frac{1}{2}Bxy \left[1 + \tanh\left(2\frac{y_d-y}{d}\right) 
                                       \right].
\end{eqnarray}

In both cases $\theta$ is set to $\pi/2$ (since neighboring leads are
perpendicular to each other), and $y_u=-y_d=(l_v + w_h)/2$.
We also define the sum of the two rotation functions as $F=F_1 + F_2$.

\begin{figure}[t]
\includegraphics{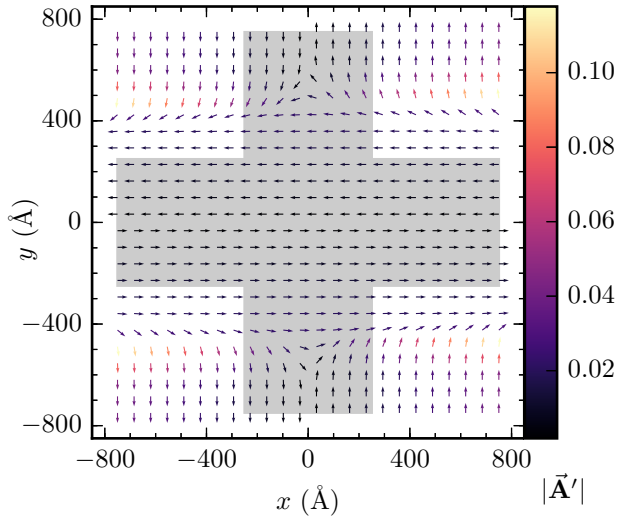}
\caption{\label{f:vector_potential} (Color online) Modified vector
    potential $\vec{A}'(x, y)$:  arrows show the direction of 
    $\vec{A}'(x, y)$, and their color represents its intensity. System shape 
    is marked by the gray area.}
\end{figure}

In order to check that the modified vector potential $\vec{A}'(x, y) =
\vec{A}(x, y) + \vec{\nabla} F(x, y)$ is properly defined, this function is
presented in Fig.~\ref{f:vector_potential}. The $\vec{A}_H$ gauge oriented
in the $x$ direction in the horizontal part of the cross transforms smoothly
to a $y$ oriented gauge $\vec{A}_V$ in the vertical part of the cross, thus
confirming the correctness of $\vec{A}'$.

Note that functions $\xi_2$ and $\xi_4$ are chosen because they are smooth, thus
guaranteeing the smoothness of the vector potential. But in a tight-binding
system, due to its discreteness, and the constant distance between the atoms, this
is not a necessary condition. The discontinuous Heaviside step function could
also be used instead. We tested this by changing the width $d$ from a value
used in all our calculations ($d=l_v/5$), to values well below the
carbon-carbon distance (equivalent to a discontinuous step function), with no
observable changes in the final results. 

The Peierls phase factor between sites $i$ and $j$, for the modified vector
potential is
%
\begin{eqnarray}
    \varphi_{ij} & = & \frac{e}{\hbar} 
                       \int_{\vec{r}_j}^{\vec{r}_i}
                       \left(
                           \vec{A}_H + \vec{\nabla} F
                       \right) 
                       d\vec{r} \nonumber \\
                 & = & \frac{e}{\hbar} 
                       \int_{\vec{r}_j}^{\vec{r}_i}
                       \vec{A}_H d \vec{r}\ +\  
                       \frac{e}{\hbar}
                       \left(F_i - F_j\right) \nonumber \\
                 & = & \varphi_{ij}^{L} + \Phi_i - \Phi_j, 
\end{eqnarray}
\noindent where $\varphi_{ij}^{L}$ is the Peierls phase factor for the
translatory invariant Landau gauge in the $x$ direction
%
\begin{equation}
    \varphi_{ij}^{L} = -\frac{e}{\hbar} B 
                        \frac{\left(y_i + y_{j} \right)}{2}
                        \left(x_i - x_j\right),
\end{equation}
as is also explained in Ref. \onlinecite{shevtsov}. Note that
$\varphi_{ij}^{L}$ does not depend on the $x$ coordinates, since differences 
$x_i - x_j$ are constant.   

Resistances in this four-terminal device are obtained using the 
Landauer-B\"uttiker formula\cite{buttiker_4term, buttiker_IBM, ferry} 
%
\begin{equation}
\label{eqR}
R_{mn,kl} = \frac{h}{2e^2}\left(T_{km}T_{ln} - T_{kn}T_{lm}\right) / D, 
\end{equation}
where  $R_{mn,kl}$ is a resistance measured when the current is injected from 
lead $m$ and collected at lead $n$, and the voltage is measured
between leads $k$ and $l$. $T_{ij}$ is the transmission function between the 
corresponding leads, while parameter $D$ is defined as 
%
\begin{equation}
\label{eqButt}
    D = \left(
            \alpha_{11} \alpha_{22}  -  \alpha_{12} \alpha_{21}
        \right)S,
\end{equation}
where
%
\begin{subequations}
\begin{eqnarray}
    \alpha_{11} & = & (T_{21} + T_{31} + T_{41}) - \nonumber \\ 
                &   & (T_{14} + T_{12})(T_{41}+T_{21})/S, \\
    \alpha_{22} & = & (T_{12} + T_{32} + T_{42}) - \nonumber \\
                &   & (T_{21} + T_{23})(T_{12}+T_{32})/S, \\
    \alpha_{12} & = & (T_{12}T_{34} - T_{14}T_{32})/S, \\
    \alpha_{21} & = & (T_{21}T_{43} - T_{41}T_{23})/S,
\end{eqnarray}
\end{subequations}
\noindent with 
%
\begin{equation}
S = T_{12} + T_{14} + T_{32} + T_{34}.
\end{equation}
%

The previous resistance formula (Eq.~\ref{eqR}) defines six different 
resistances, and when used with transmission functions at a specific 
Fermi energy $T_{ij}(E_F)$ it provides resistances for the zero temperature
case. To obtain the resistances at a nonzero temperature, the previously
calculated transmission functions need to be additionally convoluted in energy 

\begin{equation}
T'_{ij}(E_F, T) = \int_{-\infty}^{\infty} T_{ij}(E')F_T(E'-E_F)dE',  
\end{equation}
where the convolution function\cite{data} 
\begin{equation}
    F_T(E, T) = \frac{1}{4k_BT}\textrm{sech}^2\left( \frac{E - E_F}{2k_BT} \right),
\end{equation}
is the temperature dependent negative derivative $F_T(E) = -\partial f/
\partial E$ of the Fermi-Dirac distribution
\begin{equation}
    f(E) = \frac{1}{\exp[(E-E_F)/k_BT] + 1}.
\end{equation}
Since vacancy disorder introduces a considerable amount of noise in all 
calculated quantities, in some cases we perform temperature smoothing by
setting  \mbox{T = 16 K}, the temperature is considered to be zero otherwise.
In case of the averaged results, the temperature smoothing is always performed
before the averaging.

\section{Results \label{s:res}} 

\subsection{Effects of different disorder types \label{ssDisorder}}

Here we discuss the general transport effects of the three disorder types,
observable in the Hall \mbox{($R_H$ = $R_{13, 42}$)}, and the bend \mbox{($R_B
= R_{12, 43}$)} resistances. Note that actual Hall measurements are usually
performed on devices with six or more terminals, with current and voltage
probes usually set on different terminals. That is why we focus here on the
bend resistance $R_B$, and not on $R_{13,13}$, since $R_B$ should be closer to
experimentally measured $R_{xx}$. 

\begin{figure}[t]
\includegraphics{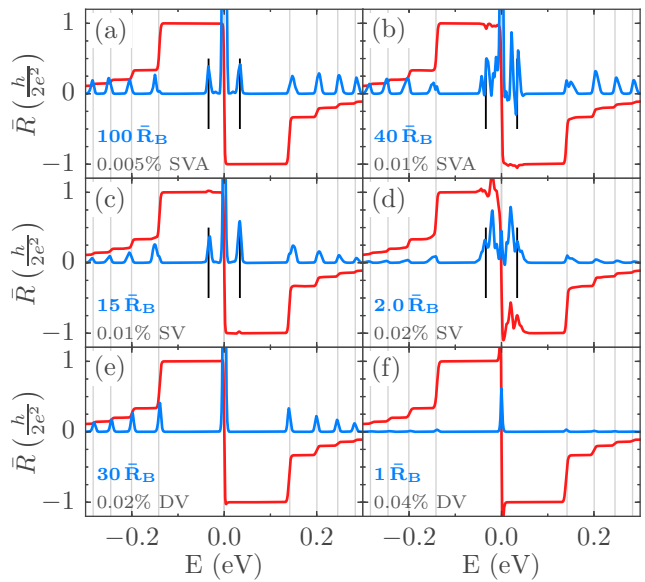}
\caption{\label{f:rbrh}
    (Color online) Average Hall (\mbox{$\bar{R}_H$ = $\bar{R}_{13, 42}$}, red 
    curves) and bend (\mbox{$\bar{R}_B = \bar{R}_{12, 43}$}, blue curves)
    resistances for various types and concentrations of vacancy disorder.
    Disorder type and concentration are shown in every subplot, in the
    lower-left corner.  Results for the bend resistance depend strongly on the
    vacancy concentration, therefore we scale $\bar{R}_B$ by multiplying it
    with a scaling coefficient, in order to present all results in the same
    range.  A scaling coefficient is presented in every subplot, above the
    vacancy concentration. Thin gray vertical lines mark the energy of Landau
    levels given by Eq.~(\ref{landau}) for $B=20$ T, while black vertical
    lines mark the position of the new peaks at $E=\pm 33.9$ meV. The 
    temperature is equal to 16 K.} 
\end{figure}

It is widely known~\cite{goerbig} that the Hall resistance in graphene
exhibits quantized plateaus 
%
\begin{equation}
    R_ H = \left(  \frac{h}{2e^2}  \right) 
           \frac{1}{1 + 2n}, 
           \ \  n = 0, \pm 1, \ldots
\end{equation}
between the energies of the Landau levels
%
\begin{equation}
    \label{landau}
    E_n  =  \text{sgn}(n)  \sqrt{2eBv_F^2\hbar |n|},
\end{equation}
where $n$ is the Landau level number, and $v_F$ is the Fermi velocity ($v_F =
3|t|a / 2\hbar\approx 10^6\ \textrm{m/s}$, with parameter \mbox{$a = 1.42$
\AA} being the carbon-carbon distance). At the steps in the Hall resistance,
the longitudinal resistance exhibits peaks. Beside these expected features,
our results for a disordered system (presented in Fig.~\ref{f:rbrh}) show some
additional features. 

 Monovacancy disorders (SV and SVA) induce two new
peaks in the bend resistance, around \mbox{$E$ = $\pm 33.9$ meV}. These peaks do not
agree with the analytic formula for Landau levels given by Eq.~(\ref{landau}).
In contrast, peaks induced by double vacancies (DV) 
appear to agree with Eq.~(\ref{landau}) (i.e.\ they correspond to the
expected Landau levels, broadened by temperature and vacancy scattering). Each
row in Fig.~\ref{f:rbrh} presents data for 
one type of vacancy disorder, for two different 
concentrations. For each vacancy type, the increase in vacancy concentration
leads to an increase in $R_B$, which can be seen in a decreasing scaling
coefficient (given in the insets of the figure). The higher the vacancy
concentration, the larger the bend resistance, and consequently the smaller
the scaling coefficient. 
For higher concentrations (Figs.~\ref{f:rbrh}(b) and (d)), two peaks in
$\bar{R}_B$ are not well defined, and $\bar{R}_H$ also slightly deviates from
the expected Hall plateaus. Although we discuss the effects of a vacancy
concentration further below, it is important to state that new peaks in
$\bar{R}_B$ occur only in a certain range of concentrations, and that above
some critical  concentration, these peaks broaden and
merge.  This critical concentration  depends on the
ratio of the average vacancy-vacancy distance and the magnetic length.
It also depends on the type of monovacancy disorder, since for SVA
disorder, the two peaks disappear for smaller concentrations (0.01\% in
Fig.~\ref{f:rbrh}(b)) as compared to SV disorder (0.02\% in
Fig.~\ref{f:rbrh}(d)).  Another interesting feature is the negative bend
resistance in Fig.~\ref{f:rbrh}(b). As explained in
Ref.~\onlinecite{buttiker_IBM} (page 321, below Eq. 13 in that reference), and
in Ref.~\onlinecite{ferry} (section 3.4.4.2 in that reference), B\"uttiker
formula for a four terminal device can produce negative non-local resistances.
This is usually the case when the second term in the numerator of
Eq.~(\ref{eqR}) is larger than the first term.  We obtain negative $R_B$ peaks
for almost all concentrations, but for low concentrations they do not appear
often (because the scattering is weak), and are not very pronounced (they
usually disappear after temperature smearing). In general, if the number of
vacancies exceeds the critical value, vacancy scattering becomes 
too strong, that no  general features exist in the low energy
region.  The bend
resistance then strongly depends on a particular vacancy distribution.

\begin{figure}[t]
\includegraphics{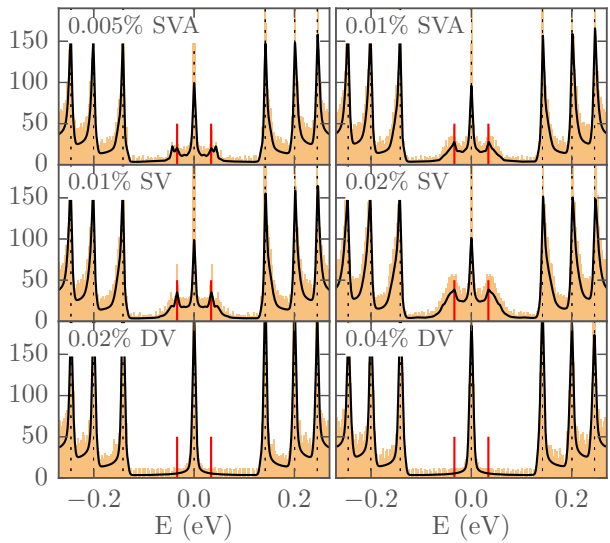}
\caption{\label{f:eigen}
    (Color online) Averaged  DOS (black curves, arbitrary 
    units) and distributions of eigenenergies in a closed system (orange 
    histograms in the background). Red lines mark the positions of the two 
    new peaks at \mbox{$E$ = $\pm33.9$ meV}, while dotted lines mark the 
    energies of Landau levels. Similarly to the resistances in 
    Fig.~\ref{f:rbrh}, for every vacancy type and concentration, DOS is 
    first smoothed and then averaged over \mbox{$N=10$} different vacancy samples. 
    In case of the eigenenergies, results for positive energies for
    \mbox{$N=10$}
    distributions are summed without smoothing or averaging, and mirrored 
    around $E=0$ axis. We used the same sets of vacancy distributions as in 
    Fig.~\ref{f:rbrh}. Magnetic field is \mbox{$B=20$ T}, and 
    \mbox{$T=16$ K}.}
\end{figure}

Another characteristic of the averaged resistance $\bar{R}_B$ is that it is
fairly symmetric  with respect to electrons and holes, whereas results for
individual distributions (used to calculate $\bar{R}_B$) are not. This means
that in general, a random  monovacancy distribution 
induces two new peaks
in the bend resistance, but the actual relative height of those two peaks
depends on a particular arrangement of vacancies. For some distributions
there is only one peak in $R_B$, at positive or negative energy, and for
some distributions there are no peaks at all (a question which we address in
subsection~\ref{ssRB}). This asymmetry between electrons and holes is
expected, since exchanging electrons for holes is equivalent to flipping the
magnetic field, which in turn is equivalent to keeping the field fixed and
flipping the system around the $z$ axis. A clean system is symmetric 
with respect to this transformation, but a disordered system is not.
After the flip, the incoming electrons see a different arrangement of
vacancies.  A vacancy distribution can be
constructed to be symmetric with this flip transformation, in which case all
results would also be electron-hole symmetric. This asymmetry between
electrons and holes occurs only for a fixed field orientation ($R_B(E,
B)\neq R_B(-E, B)$), and should not be confused with the case when both
magnetic field and Fermi energy change sign. Results for electrons and holes
are then symmetric: $R_B(E, B)=R_B(-E,-B)$.

Results for the averaged total density of states (DOS) and distributions of
eigenlevels in a closed system, presented in Fig.~\ref{f:eigen}, are obtained
for the same set of vacancy distributions as  in Fig.~\ref{f:rbrh}, and
they exhibit similar effects to those seen in the bend resistance in
Fig.~\ref{f:rbrh}. Here, as in Fig.~\ref{f:rbrh},  
monovacancy distributions induce two new broad peaks in the total DOS
(around \mbox{$E$ = $\pm$33.9} meV, marked with red lines in Fig.~\ref{f:eigen}),
while double-vacancy distributions appear only to broaden
the DOS around the expected Landau levels. We show below that the
energy of these broadened peaks \mbox{($E=\pm 33.9$ meV)} corresponds to an
energy of a  monovacancy localized state.
Similar behaviour is seen in distributions of eigenlevels in a closed system
(a Hall bar detached from the leads). Usually, the eigenlevels of a closed
system in a high magnetic field tend to cluster around the energies of the
Landau levels. Here we plot histograms (orange areas in Fig.~\ref{f:eigen})
showing how many eigenlevels occupy a narrow energy range around each energy,
and these plots also show two distributions around \mbox{$E$ = $\pm 33.9$
meV}. 

\begin{figure}[t]
\includegraphics{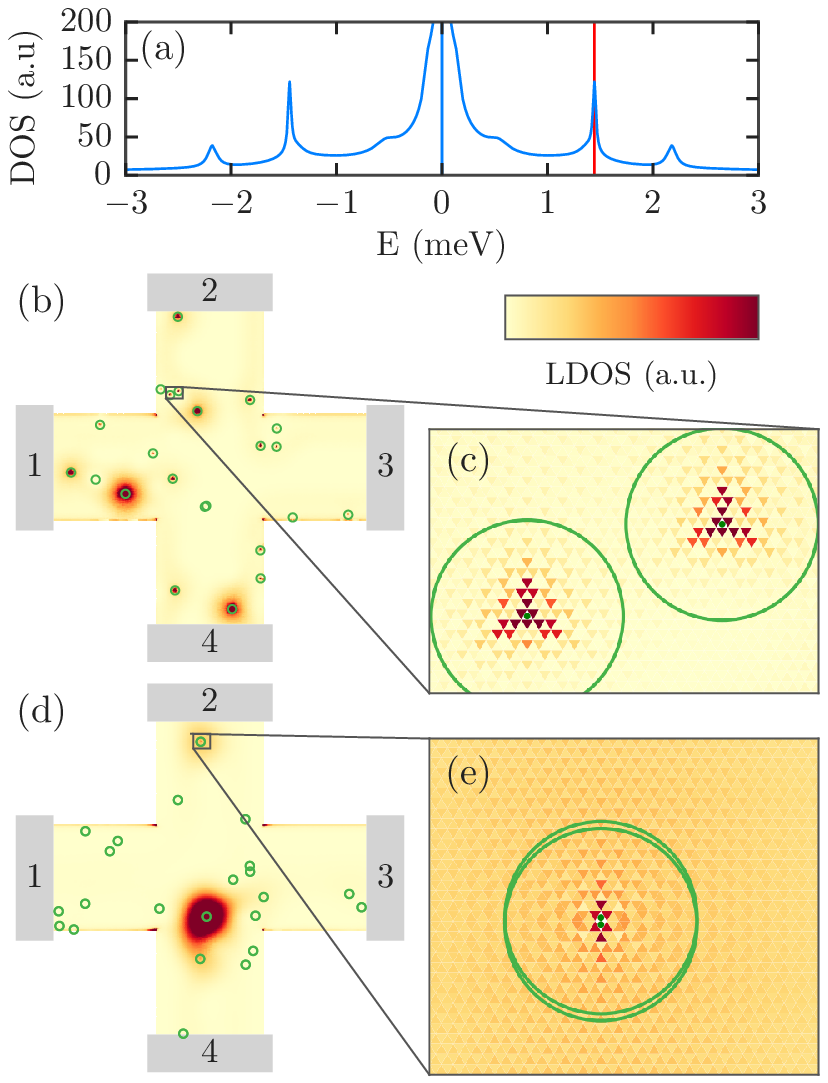}
\caption{\label{f:ldos}
    (Color online) (a) DOS for a single DV distribution
    \mbox{($n=0.01$\%, $B=14$ T)} (b) LDOS for a single SVA distribution
    \mbox{($n=0.005$\%, $B=20$ T)} at the vacancy localization energy
    \mbox{($E=33.9$ meV)}. (d) LDOS for a single DV distribution 
    \mbox{($n=0.01$\%, $B=14$ T)} at the localization energy 
    (\mbox{$E=1.45$ meV}, marked with red vertical line in (a)). (c) and (e)
    Zoom around particular vacancies in (b) and (d). Vacancies are marked with
    green circles and green dots in the center. Temperature is set to
    \mbox{$T=0$ K}.} 
\end{figure}

\begin{figure*}[t]
\includegraphics{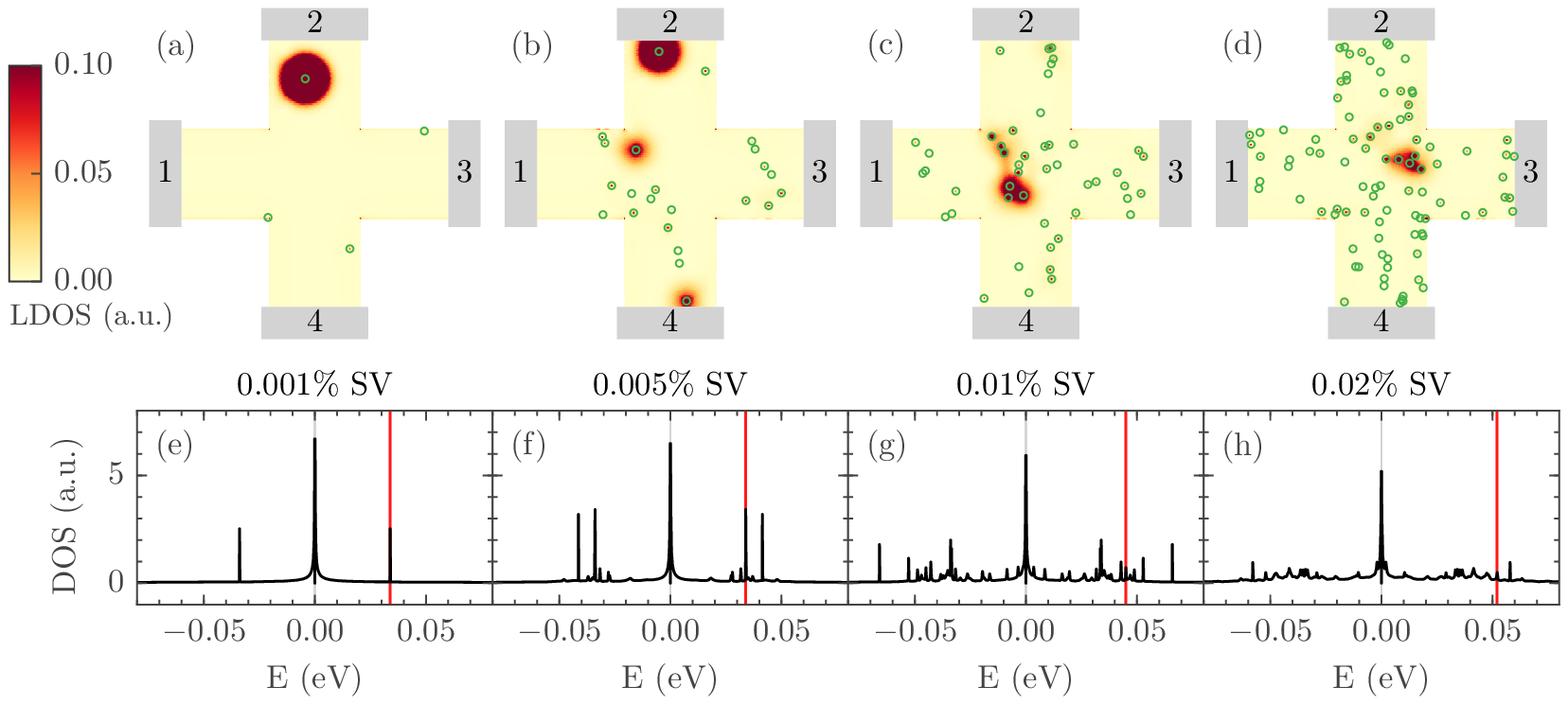}
\caption{\label{f:density}
    (a-d) LDOS for SV disorder type, for different vacancy concentrations: 
    0.001\%, 0.005\%, 0.01\%, and 0.02\%, and at different energies:
    \mbox{$33.9$ meV}, \mbox{$33.9$ meV}, \mbox{$45$ meV}, and \mbox{$52$
    meV}, respectively. Vacancy positions are marked with green circles.
    \mbox{(e-h)} DOS for SV disorder type, for the corresponding
    concentrations. Energies at which we calculated LDOS in (a-d), are marked
    with red vertical lines in (e-h). Magnetic field is \mbox{$B=20$ T}, and
    \mbox{$T=0$ K}.} 
\end{figure*}

According to Refs.~\onlinecite{pereira_vacancies, leconte_prb_2016},
divacancies in graphene should also induce new states in the Landau spectrum.
Our results for DV distributions appear to contradict those of these two
references. However, a higher resolution DOS plot around the zeroth Landau
level (shown in Fig.~\ref{f:ldos}(a)) for one particular DV distribution,
reveals additional DOS peaks. These peaks are positioned only few meV away from the
LLs, and that is why they were not well distinguishable from the LLs
in the previous results.  This suggests that additional peaks coming from
the divacancies would be harder to observe experimentally, since they would
be usually smeared by temperature. 

Previous experiments on graphene in high magnetic fields~\cite{li.prl.2009}
reported splitting of the zero Landau level, which was attributed to the
breaking of the sublattice symmetry. One of the possible explanations of the
new DOS peaks  is that they correspond to the occurrence of new states,
localized  in areas close to the vacancies. The unsplit zeroth Landau level is
still present in the DOS of the whole device (for all disorder types), since
it originates from local density of states (LDOS) in areas which are vacancy
free. This connection between the new DOS peaks and  the vacancy localized
states becomes apparent if we look at the LDOS at one of the two peak
energies.  A LDOS at one of the two new peaks, for one particular SVA
distribution, is shown in Fig.~\ref{f:ldos}(b). LDOS is highly localized
around the vacancies (marked with green circles). A zoom in
Fig.~\ref{f:ldos}(c) shows states localized mostly on one sublattice, which
could be connected with the breaking of sublattice symmetry.  These states
that are localized around single vacancies are the origin of the two new peaks
in $R_B$ and {DOS}. Similarly, in Figs.~\ref{f:ldos}(d) and (e) we show LDOS for
one particular DV distribution, at energy of one of the new peaks (marked with
red line in Fig.~\ref{f:ldos}(a)). Divacancies also induce localized states,
but these states are localized equally on both sublattices, since divacancies
do not break the sublattice symmetry.  For other energies (when there is no
localization), divacancies  act similarly to graphene structural armchair
edges, namely LDOS spreads in areas between them, as if they repel it. Similar
behaviour was observed in Ref.~\onlinecite{leconte_prb_2016}.

\subsection{Changing vacancy concentration\label{ssConcentration}}

As stated previously, all these results depend strongly on the vacancy
concentration. To illustrate this, we present in Fig.~\ref{f:density} how DOS
and LDOS change with increasing number of vacancies, belonging to a SV
disorder type. Here, we show results for specific vacancy distributions
without any temperature smoothing or averaging. For low concentrations
(Figs.~\ref{f:density}(a) and~\ref{f:density}(e)) DOS shows two well-defined
peaks at $\pm$33.9 meV, which correspond to one state, localized around one
monovacancy. Other vacancies in Fig.~\ref{f:density}(a) are very close to the
system edges, and localization on them is very weak. These results explain why
smoothed and averaged $R_B$ and DOS exhibit strong peaks around \mbox{$\pm
33.9$ meV}, because this corresponds to the energy of a state localized around
one  isolated monovacancy.  The localization happens at this specific energy
only if a vacancy is in the bulk and sufficiently away from the system edges,
but also far from the other vacancies, which is satisfied only for low
concentrations. With increasing concentration, the average distance between
the vacancies decreases, and vacancies start to ``see'' each other, meaning
that they start to influence the formation of localized states on their
neighbours. This is demonstrated in Figs.~\ref{f:density}(b) 
and~\ref{f:density}(f), where several peaks appear in the total {DOS}. 
However, the
peak at the  monovacancy localization energy ($\pm 33.9$ meV) is still well
defined. This is because there is still one well isolated monovacancy  in the
upper arm of the cross (see Fig.~\ref{f:density}(b)). For even higher
concentrations, there are no longer well isolated vacancies, and therefore
there is no well defined localization energy. Instead, the monovacancies start
to form something which resembles bond states. In a vague analogy with atoms
and molecules, these bond states correspond to groups of vacancies which are
sufficiently close to each other, so that localization occur over the whole
group, and not on separated, individual vacancies. This bonding, shown in
Figs.~\ref{f:density}(c) and~\ref{f:density}(d), is responsible for spreading
of the localization energy, and consequently for broadening of the new peaks
in $R_B$ and DOS.

\subsection{Changing the magnetic field\label{ssBfield}}

\begin{figure}[h]
\includegraphics{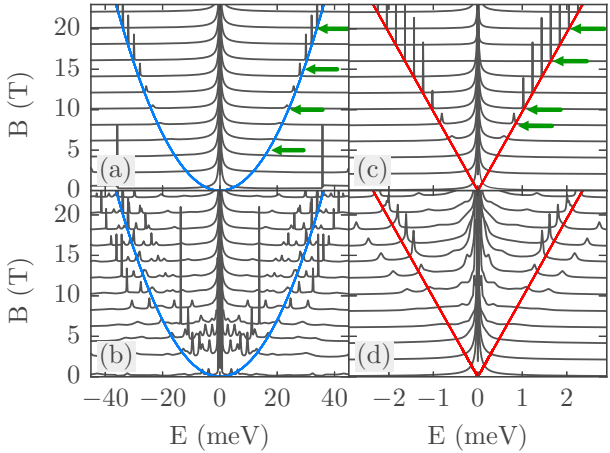}
\caption{\label{f:e_vs_b}
    (Color online) DOS for different values of magnetic field and different
    disorder distributions: (a) \mbox{$n=0.001$\%} SV distribution, (b)
    \mbox{$n=0.005$\%} SV distribution, (c) single divacancy located at the
    center of the system, (d) \mbox{$n=0.01$\%} DV distribution. The two
    distributions of monovacancies are the same as those used in
    Figs.~\ref{f:density}(a) and~\ref{f:density}(b), respectively. In all four
    cases \mbox{$T=0$ K}.  The green arrows mark \mbox{($E, B$)} points at
    which we study LDOS in Figs.~\ref{figLvsBa},~\ref{figLvsBb}, 
    and~\ref{figLvsB_DV}.} 
\end{figure}

In this part, we investigate how these vacancy localized states behave when we
change the magnetic field $B$. In Figs.~\ref{f:e_vs_b}(a)
and~\ref{f:e_vs_b}(b) we show that the localization energy  for monovacancies
scales with the square root of the magnetic field $E\sim \pm\sqrt{B}$,
similarly to the relativistic Landau levels.  The blue curves in
Figs.~\ref{f:e_vs_b}(a) and~\ref{f:e_vs_b}(b) show the parabolic function
\mbox{$B=\alpha E^2$}, where parameter \mbox{$\alpha=17500$ T/(eV)$^2$} is set
to fit the peak positions. This dependence can also be expressed as
\mbox{$E=\pm \sqrt{\gamma 2 e v_F^2 \hbar B}$}, where
\mbox{$\gamma\approx0.057$}.  It is important to note that
Fig.~\ref{f:e_vs_b}(a) presents results for the same vacancy distribution as
in Fig.~\ref{f:density}(a), with only one monovacancy  capable of sustaining
the localized states.  The DOS in this case exhibits two narrow peaks at
positive and negative localization energy. For weak fields \mbox{($B<5$ T)}
these peaks are almost unobservable, whereas for stronger fields they become
better and better defined in energy. Beside these two localization peaks,
Fig.~\ref{f:e_vs_b}(a) shows some additional peaks for \mbox{$B=0$ T} 
(e.g.~two peaks at approximately  $\pm35$~meV). According to Pereira et
al.~[\onlinecite{pereira_prb}], localization of electrons on vacancies also
occurs for \mbox{$B=0$ T}, but localization energy is then equal to zero, and
therefore these extra peaks should not be connected with the localized states.
Indeed, a closer study reveals that these peaks originate from new modes
opening in the leads, and can be predicted by calculating the lead minimal
subband energies. 

For larger concentration of monovacancies
(Fig.~\ref{f:e_vs_b}(b)), the localization energy is not well defined, and the
two narrow DOS peaks from Fig.~\ref{f:e_vs_b}(a) split into two distributions
of peaks. As we explained in the previous subsection, this is mainly due to a
decrease of the average vacancy-vacancy distance, and is thus due to an
increase of the interference between vacancies, resulting in the formation of
bond localized states. Since localization radius around a  monovacancy is 
inversely proportional to the square root of the magnetic field (as we show
below), the field strength determines how far a single vacancy actually
``sees'' its surroundings, i.e.~it determines the bond length of previously
described bond states. Because this length changes with magnetic field,
various groups of vacancies bond together at different field strengths, and
the two distributions of DOS peaks in Fig.~\ref{f:e_vs_b}(b) evolve quite
unpredictably with $B$. However, the average energies of the two distributions
still tend to follow the parabolic $B$ dependence, as is apparent from the
graph. Thus for extremely large fields, the localization would be so strong
that the bond length will go below the average vacancy-vacancy distance, and
the vacancies would no longer ``see'' each other. All these separate DOS peaks
would then converge to a single energy, equal to that of an isolated
monovacancy. 

\begin{figure}[t]
\includegraphics[]{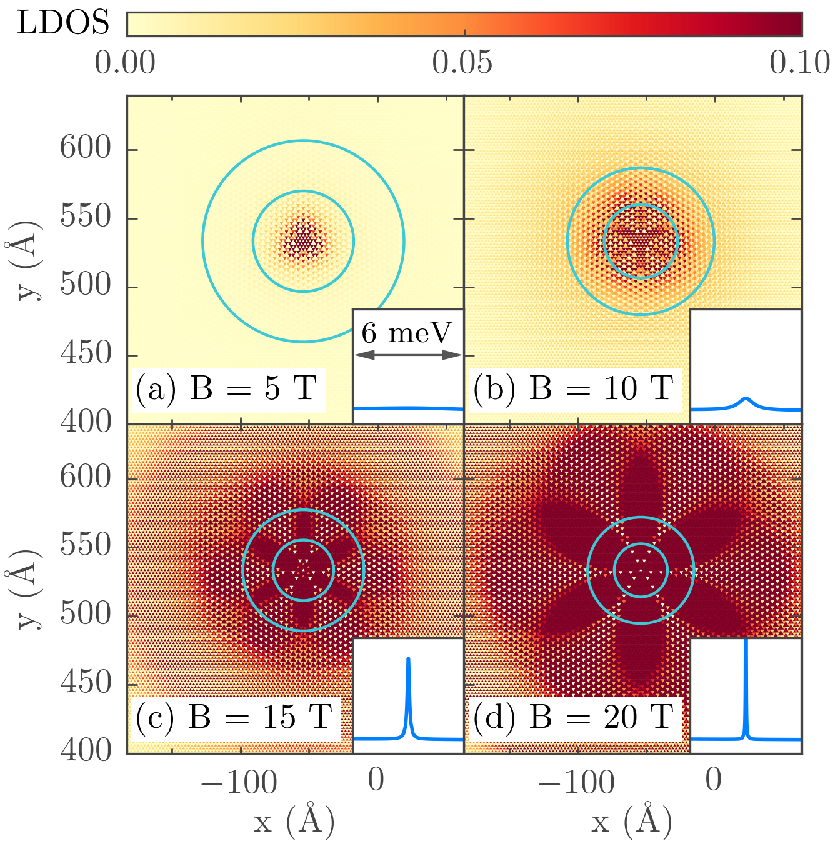}
\caption{\label{figLvsBa} 
    (Color online) The evolution of LDOS around a single SV vacancy for
    \mbox{$(E, B)$} values lying on the parabola in Fig.~\ref{f:e_vs_b}(a)
    (marked by the green arrows). The magnetic field strengths are $5$~T,
    $10$~T, $15$~T, and $20$~T, and the corresponding energies and cyclotron
    radii are (a) \mbox{$E=16$ meV}, \mbox{$R_c = 36.7$ \AA}, (b)
    \mbox{$E=23.4$ meV}, \mbox{$R_c=26.8$ \AA}, (c) \mbox{E = 29 meV},
    \mbox{$R_c$ = 22.1 \AA}, and (d) \mbox{$E=33.9$ meV}, \mbox{$R_c=19.4$
    \AA}. The two circles with radii $R_c$ and $2R_c$ in each inset are
    centered at the vacancy site. Insets in the lower-right corners show the
    total DOS in a \mbox{$6$ meV} energy range around the localization
    energy.} 
\end{figure}

\begin{figure}[tbh]
\includegraphics[]{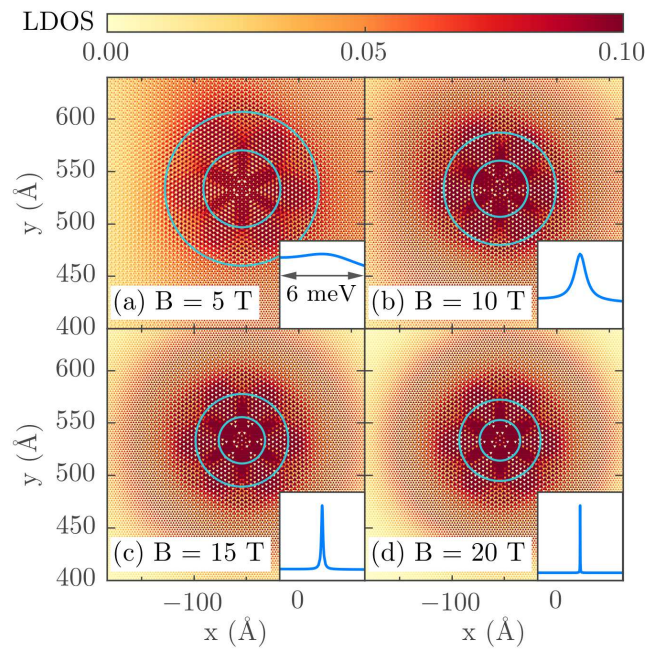}
\caption{\label{figLvsBb} 
    (Color online) Same as Fig.~\ref{figLvsBa} but now showing normalized
    LDOS, where LDOS in each subplot is divided by a maximum LDOS value for
    that subplot. The lower-right insets (showing the total DOS around the
    energy of a localized state) are also scaled, so that the DOS peak 
    maximum is equal to one.} 
\end{figure}

\begin{figure}[tbh]
\includegraphics[]{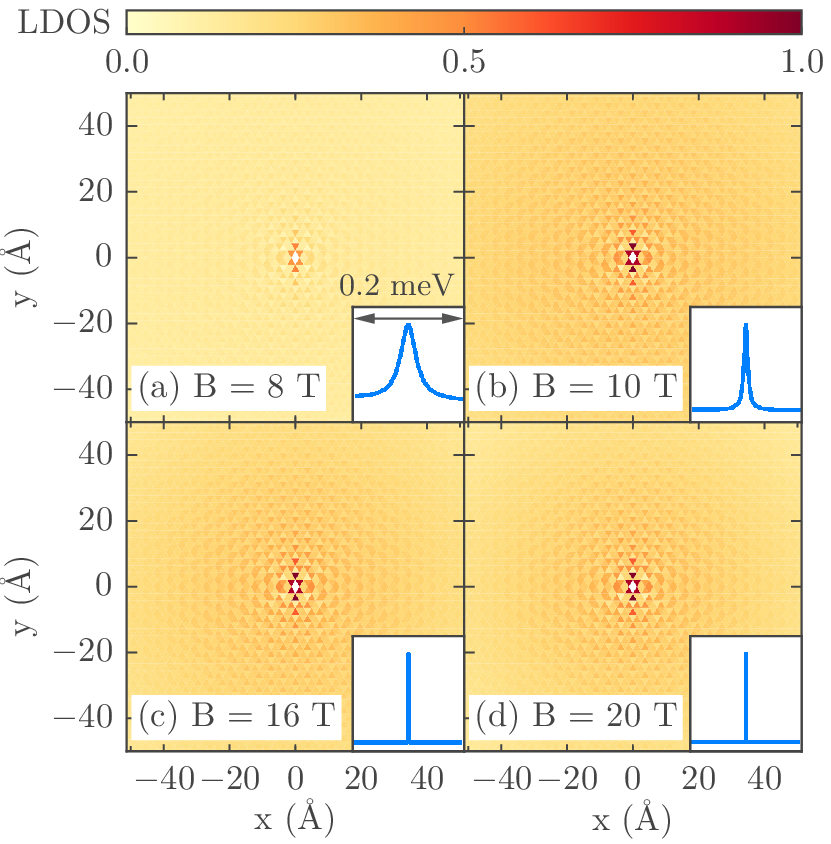}
\caption{\label{figLvsB_DV} (Color online) Same as Fig.~\ref{figLvsBb} but
now showing normalized LDOS and DOS for a single divacancy located at the
center of the system. The (E, B) points at which we calculated LDOS and DOS
are marked with green arrows in Fig.~\ref{f:e_vs_b}(c).} 
\end{figure}

Scaling of the localization energy with magnetic field is different
for divacancies. As Figs.~\ref{f:e_vs_b}(c) and~\ref{f:e_vs_b}(d) show, the
localization energy for divacancies scales linearly with the magnetic field.
The red lines in these two figures mark the linear dependence 
\mbox{$B=\beta E$}, where \mbox{$\beta\approx 9700$ T/eV}.
Contrary to monovacancies where the bond states evolve rather unpredictably 
with magnetic field, the bond states of divacancies evolve very predictably 
with the field. The two DOS peaks at positive energies, and the two peaks 
at negative energies in Fig.~\ref{f:e_vs_b}(d) move proportionally to the 
magnetic field. There are no additional peaks which would correspond to
different bonding of divacancies. These results suggest that bonding of 
divacancies is weaker, when compared to monovacancies. One of the possible 
reasons for this weaker bonding might be the constant localization length for 
divacancies, which we discuss below.

Reference~\onlinecite{pereira_vacancies} also studied the $E(B)$
dependence of the new (localized) states, and for both mono- and divacancies
found it to be neither linear, nor parabolic. However, the lowest field
considered in that reference (beside \mbox{$B=0$ T}) was around \mbox{$300$
T}, therefore our results can be understood as a low field limit of those
presented in Ref.~\onlinecite{pereira_vacancies}.

As we stated previously, the localization radius for monovacancies $r_L$ is
inversely proportional to the square root of the magnetic field. It is also
proportional to the cyclotron radius $r_L \sim R_c = E/(ev_FB)$, and since
$E\sim \sqrt{B}$, then $r_L \sim 1/\sqrt{B}$. To demonstrate this, in
Fig.~\ref{figLvsBa} we follow how LDOS around an isolated monovacancy evolves
as we increase the magnetic field. In other words, we follow the localized
state along the $\alpha E^2$ parabola in Fig.~\ref{f:e_vs_b}(a). A first look
at Fig.~\ref{figLvsBa} suggests that localization radius is not proportional
to the cyclotron radius $R_c$. While $R_c$ decreases with rising magnetic
field, the localization radius appears to increase, and LDOS forms intricate
flower-like patterns. The answer to this contradiction lies in the lower-right
insets in Fig.~\ref{figLvsBa}, which show the total DOS around the
localization energy. For stronger fields, the localized state is better
defined in energy, and therefore the total DOS is larger. In order to properly
compare these four cases, we need to normalize the LDOS in each subplot. This
is done in Fig.~\ref{figLvsBb}, where each LDOS distribution is divided by its
maximal value. Now, with these normalized results, the localization radius
scales proportionally to the cyclotron radius $R_c$ as is intuitively
expected. The scaled results also point  to another interesting feature. We
stated earlier that the LDOS around a monovacancy is localized mostly on one
sublattice, opposite to that of the vacancy. However, the scaled results show
that at the localization energy, LDOS around a monovacancy spreads over  both
sublattices.  Nonzero LDOS on the vacancy sublattice is located mostly in the
symmetric, flower-like area. Outside of this area, states are still localized
only on one sublattice. LDOS is also $C_{3v}$ symmetric, which can be
connected with the underlying $C_{3v}$ (structural) lattice symmetry.

A similar LDOS comparison, but for an isolated divacancy, is presented in
Fig.~\ref{figLvsB_DV}. Contrary to monovacancies, a divacancy localization
length does not change significantly with magnetic field. This can be
understood just based on the linear $E(B)$ dependence of the divacancy
localization energy. If we assume that localization length is still
proportional to the cyclotron radius, then since \mbox{$E = B / \beta$}, it
follows that $r_L\sim R_c=E /(ev_F B)=1/(\beta ev_F)$. LDOS around a divacancy
is $C_{2v}$ symmetric, which could be also connected with the underlying
lattice symmetry. Contrary to monovacancies, divacancies preserve the
sublattice symmetry, and this is the origin of the different behaviour of
these two disorder types.


\begin{figure}[t]
\includegraphics[width=0.45\textwidth]{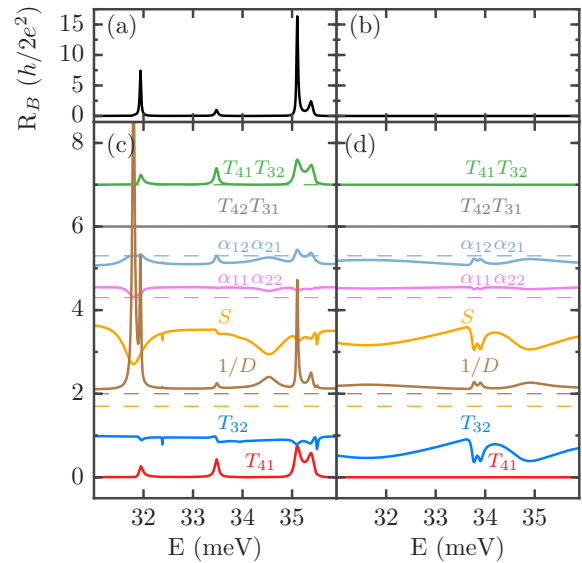}
    \caption{(Color online)\label{figRb} Decomposition of the bend resistance
    \mbox{$R_B = \left(T_{41}T_{32} - T_{42}T_{31}\right)/D$}, where
    \mbox{$D=(\alpha_{11}\alpha_{22} - \alpha_{12}\alpha_{21})S$.} Results for
    two SV distributions \mbox{($n = 0.01$ \%)} in (a) and (b) are decomposed
    in (c) and (d), respectively. Most of the terms in (c) and (d) are
    vertically displaced, with dashed lines marking the corresponding
    positions of the zero axes. Magnetic field is \mbox{$B=20$ T}, and \mbox{T
    = 0 K}.  } 
\end{figure}

\subsection{Decomposition of $R_B$ and the current density \label{ssRB}}

When discussing results for the averaged bend resistances $\bar{R}_B$
in~\ref{ssDisorder}, we mentioned that although the averaged results appear to
be symmetric for electrons and holes, the results for individual distributions
are not, and for some distributions there are no new peaks in $R_B$. In this
subsection we study why this is the case.  We compare bend resistances for two
specific  monovacancy distributions: one for which there are new peaks in
$R_B$, and one for which there are not. In order to understand how these peaks
come into existence from the different transmission terms in the $S$ matrix,
we decompose $R_B$ on its constituent parts, according to the
Landauer-B\"uttiker (LB) formula.~\cite{buttiker_4term} Results are presented
in Fig.~\ref{figRb}. Here we focus only on a narrow energy range where these
new peaks in $R_B$ appear. Analysis of the main LB terms in
Fig.~\ref{figRb}(c) reveals that only one term ($T_{41}T_{32}$, green curve)
is responsible for the appearance of the $R_B$ peaks. The other term in the
numerator ($T_{42}T_{31}$, gray line) is always equal to zero. A further
decomposition of the first term ($T_{41}T_{32}$, green curve) shows that one
transmission function ($T_{32}$, blue curve) is very close to unity, and that
only $T_{41}$ (red curve) dictates where the new $R_B$ peak appear. Only when
this transmission ($T_{41}$) is nonzero, we have peaks in $R_B$.  Therefore,
to a first approximation, we can say that $R_B$ is proportional to modulated
$T_{41}$. One might argue that $T_{32}$ is also important, but since $B$ is
perpendicular, $T_{32}$ will always be close to unity in this energy range,
because of the edge states that go from the second to the third lead. This
$R_B$-$T_{41}$ connection is also confirmed in Figs~\ref{figRb}(b) 
and~\ref{figRb}(d), where both main LB terms in the numerator are equal to 
zero, as well as $T_{41}$, and thus $R_B$ is also equal to zero. 

\begin{figure}[th]
\includegraphics{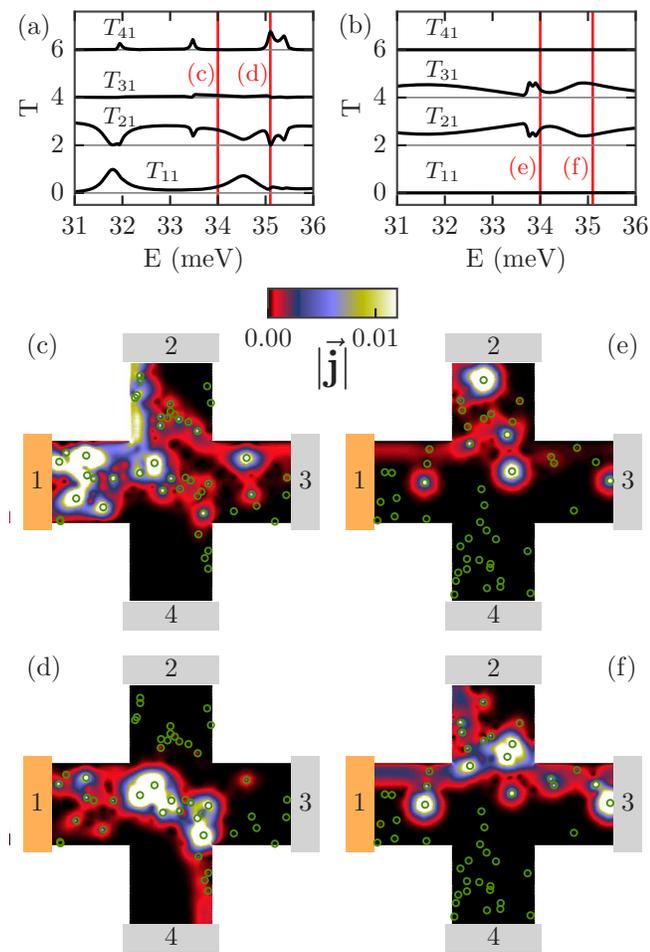}
    \caption{\label{figCurrent} (Color online) (a-b) Electron transmission
    functions from the first lead, for two specific disorder distributions
    (the same two distributions as used in Figs.~\ref{figRb}(a) 
    and~\ref{figRb}(b), respectively). The transmission functions are 
    vertically displaced by $\Delta T = 2$ for clarity. \mbox{(c-f)} Current
    densities at energies marked in (a) and (b) by vertical red lines.
    Magnetic field is \mbox{$B=20$ T}, and temperature is \mbox{$T$ = 0 K}.} 
\end{figure}

The only way to understand why for some vacancy distributions the particular
$T_{41}$ transmission is equal to zero, and for some it is not, is to
investigate how electron current flows in the presence of vacancy disorder.
This is presented in Fig.~\ref{figCurrent}, for the same two SV distributions
as those used in Fig.~\ref{figRb}. In a clean system without vacancies, and
with a perpendicular magnetic field, all current from the first lead would go
to the second lead because of the current carrying edge states.  This is
mostly what we see in both cases (Figs.~\ref{figCurrent}(a) 
and~\ref{figCurrent}(b)) where the $T_{21}$ term is the most dominant when
compared with the other transmission functions. This is also visible in
Fig.~\ref{figCurrent}(c), where most of the current from the first lead
travels to the second lead along the edges. The two vacancy distributions
differ in the way they scatter this edge current from the first to the second
lead. The first distribution (Fig~\ref{figCurrent}(a)) is causing more
backscattering ($T_{11}$), and scattering to the fourth lead ($T_{41}$),
whereas the second distribution is causing more scattering to the third lead
($T_{31}$ in Fig~\ref{figCurrent}(b)).  Where this edge current is diverted
depends mostly on a particular arrangement of vacancies, since current flow is
pinned by the vacancies. For example $T_{41}$, and consequently $R_B$,
exhibits narrow peaks because of a particular arrangement of vacancies in the
central part of the cross. As shown in Fig.~\ref{figCurrent}(d), the current
starts to flow around these vacancies, and it is diverted to the fourth lead.
In a similar way, the current flow in Figs.~\ref{figCurrent}(e) 
and~\ref{figCurrent}(f) is also pinned by the vacancies, and diverted to the
third lead. 

Reference~\onlinecite{data} (and particularly chapter IV in this reference)
gives a valuable explanation of the QHE in terms of the electron propagation
along the sample edges.  According to this reference, the rise of longitudinal
resistance (for Fermi energies coinciding with the Landau levels) occurs due
to the existence of states in the interior of the sample. These bulk states
connect the otherwise separated edges channels, and give rise to their
backscattering, and this backscattering manifests in a nonzero longitudinal
resistance. The existence of vacancy localized states in our system, with
energies in-between the Landau levels, leads to the expected LL broadening.
Additionally, these vacancy states can provide a narrow pathway between  the
channels propagating along the opposite edges of the system. In our particular
setup, the nonzero $T_{41}$ term is due to a backscattering between a channel
going from the 1st to the 2nd lead (1$\rightarrow$2), and the one going from
the 3rd to the 4th lead (3$\rightarrow$4). This edge state scattering is
responsible for the nonzero bend resistance.

Similar analysis can also explain the asymmetry between the resistance results
for electrons and holes (when the field direction is fixed). In a clean sample
with no vacancies, \mbox{$T_{21}=1$} and \mbox{$T_{41}=0$} for electrons,
while \mbox{$T_{21}=0$} and \mbox{$T_{41}=1$} for holes.  Also
\mbox{$T_{32}=1$} for electrons, while \mbox{$T_{32}=0$} for holes. We already
showed that the first term in the B\"uttiker formula (term $T_{41}T_{32}$ in
Eq.~(\ref{eqR})) determines the bend resistance. For electrons this term
depends mostly on $T_{41}$, since \mbox{$T_{32}=1$}. For holes, on the other
hand, it depends on $T_{32}$, since $T_{41}=1$.  Because $T_{32}(-E, B)\neq
T_{41}(E, B)$, the bend resistance in a disordered system is not the same for
electrons and holes. The bend resistance becomes equal only if we additionally
change the magnetic field direction (from $\vec{B}$ to $-\vec{B}$) when we
switch from electrons to holes.

In summary, although the two new peaks in $R_B$ should in general appear at
the vacancy localization energy, they are very sensitive to a particular
distribution of vacancies. The vacancies significantly disrupt and divert the
current flow. However, if not in $R_B$, this
current guiding will probably manifest itself in measurements of some other
non-local resistance.

\subsection{NNN interaction\label{ssNNN}}

\begin{figure}[tb]
\includegraphics[]{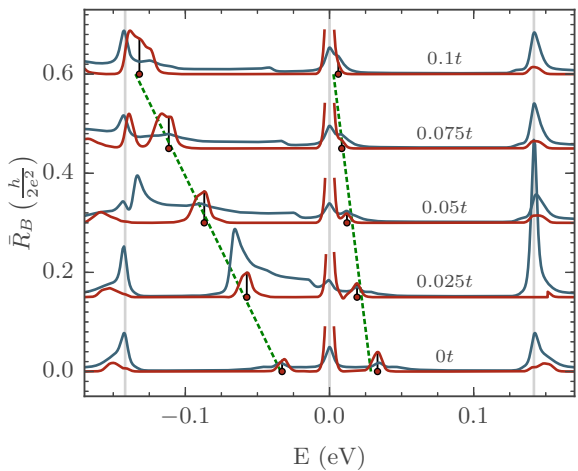}
    \caption{\label{f:NNN} (Color online) Smoothed and averaged bend
    resistances ($\bar{R}_B$, red curves), and total DOS (blue curves) for 
    increasing value of the next-nearest neighbour hopping energy $t'$. All
    results are obtained for the SV disorder type (\mbox{$n=0.01$\%}), for
    \mbox{$N=10$} different disorder distributions. Magnetic field is
    \mbox{$B=20$ T}, and \mbox{$T=16$ K}. Green, dashed lines are linear fits
    of the $\bar{R}_B$ peak energy versus the NNN hopping energy $t'$. Results
    for $t' \neq 0$ are displaced horizontally by \mbox{$\Delta E = -3|t'|$} 
    in order to align the Landau levels.  The $R_B$ 
    peaks at the zeroth Landau level are cut-off above 0.09 $h/(2e^2)$ for 
    clarity.}
\end{figure}

In this section we study the effects of a nonzero hopping between the second
nearest neighbours ($t' \neq 0$). Figure~\ref{f:NNN} shows the averaged
results for the SV disorder type, for increasing value of the next-nearest
neighbour (NNN) hopping. According to Pereira et al.
(Refs.~\onlinecite{pereira_prl, pereira_prb}), for the $B = 0$ case, there are
vacancy localized states even when $t' \neq 0$. Although the NNN hopping
breaks the electron-hole symmetry, the localized states are still preserved.
Here, we study the non-zero magnetic field case, and we still observe
localization peaks.  Breaking of the e-h symmetry leads to a displacement of
the two peaks, and this displacement (as we show in Fig.~\ref{f:NNN}) depends
linearly on the NNN hopping energy $t'$. One of the peaks moves toward the
$n=-1$ Landau level, whereas the other moves to the zeroth Landau level.
Although the two new peaks are clearly visible in the bend resistance, they
are not so distinguishable in the {DOS}. {DOS} exhibits considerable 
broadening, and the two peaks are barely visible after temperature smoothing.
A closer look in the LDOS for $t' \neq 0$ (not shown) reveals a strong
localization on the horizontal, zigzag edges. This edge localization causes
this wide background in DOS and masks the narrow vacancy localization peaks.

The linear energy dependence of the new peaks can be further explained if
compared with the zero field results of Ref.~\onlinecite{pereira_prb}. The
introduction of a nonzero NNN hopping shifts the whole Landau spectrum by
$\Delta E=3|t'|$. According to Ref.~\onlinecite{pereira_prb}, the shift of the
{\it zero mode}\ is less than $\Delta E$, and proportional to $t'$. If we
assume that the two localized states that we obtain originate from this
shifted zero mode, than we expect them to also shift linearly and follow the
zero mode. On the other hand, the two linear coefficients in Fig.~\ref{f:NNN}
are different (the distance between the peaks increases with $t'$). If we
assume that the parabolic $E(B)$ dependence is preserved, than we can conclude
that $t'$ also modifies the scaling coefficient $\alpha$, which is then a
linear function of $t'$. 

\section{Conclusions \label{s:end}}
To conclude, we  studied electron transport in graphene Hall bars in
quantizing magnetic fields in the presence of three different types of vacancy
disorder.   All three types of vacancy disorder induce new states in the
relativistic Landau spectrum, but these states behave differently depending on
the disorder type. The new states, localized around monovacancies,  are
indirectly observable in the bend resistance and in the total DOS, but only
for vacancy concentrations below a critical concentration.  These states are
localized mostly only on one sublattice, but at the localization energy they
spread on both sublattices in $C_{3v}$ symmetric, flower-like patterns.
Another interesting feature is the different behaviour of the two monovacancy
distribution types. SVA disorder, although inducing approximately equal number
of states as SV disorder (compare DOSs in Fig.~\ref{f:eigen} for these two
disorder types for \mbox{$n = 0.01$\%}) creates considerable different results
in the bend resistance. The origin of these differences is not known, and
requires further study. We speculate that these differences might come from
different current flow patterns around different types of vacancy pairs.  For
example, Ref.~\onlinecite{pereira_vacancies} showed that vacancy coupling does
not depend on their type, however, we showed that these states have a certain
symmetry, therefore the coupling strength will also depend on direction, and
not only on distance.  Divacancies also cause localization, but for fields
that we consider, their localization energies are much closer to those of the
relativistic LLs, which makes them harder to observe experimentally.  Since
they do not break the sublattice symmetry, they are usually $C_{2v}$
symmetric, and they have a constant localisation length. 

Depending on the ratio between the average vacancy-vacancy distance (which
depends on the vacancy concentration) and the field strength, localized states
around several monovacancies can bond together, forming localized bond states.
These bond states have a localization energy different from that of an
isolated  monovacancy, but on average they spread equally around this energy.
The localization energy around a single monovacancy is proportional to the
square root of the magnetic field, while the localization radius (and
consequently the possible radius of the bond states) scales with the cyclotron
radius. The behaviour of divacancies is different.  Their localization energy
scales linearly with the field, and their localization length is independent
of the field. Based on this, whether they form bond states depends solely on
their mutual distance, and not on the field strength. 

A decomposition of the bend resistance reveals that only one transmission
function ($T_{41}$) is responsible for the appearance of additional peaks in
$R_B$, which we additionally connect with the vacancy guided current flow
inside the system. All these results are slightly modified when a next-nearest
neighbor interaction is included, and the symmetry between electrons and holes
is broken.

\section{Acknowledgements} 

This work was supported by the Methusalem programme of the Flemish government.


%

\end{document}